\documentclass[traditabstract, longauth]{aa} % for the abstract without structuration 
\usepackage{txfonts}
\usepackage{graphicx}
\usepackage{color}
\usepackage{natbib}
\bibpunct{(}{)}{;}{a}{}{,}
\def\cmmt{\rm {cm^{-2}}}

\def\s-1{\rm {s^{-1}}}

\def\HC3N{HC$_3$N}

\def\kms{\hbox{${\rm km\,s}^{-1}$}}
\def\msun{M$_{\odot}$}
\def\lsun{L$_{\odot}$}

\newcommand{\asec}{\mbox{$''$}}

\usepackage{url}

\begin{document}
 \title{A precessing molecular jet signaling an obscured, growing supermassive black hole in NGC1377?\thanks{Based on observations
carried out with the ALMA Interferometer. ALMA is a partnership of ESO (representing its member states),
NSF (USA) and NINS (Japan), together with NRC (Canada) and NSC and ASIAA (Taiwan),
in cooperation with the Republic of Chile. The Joint ALMA Observatory is operated by ESO, AUI/NRAO and NAOJ.}}

\author{S. Aalto
          \inst{1}
          \and
	  F. Costagliola\inst{1}
     \and
        S. Muller\inst{1}
          \and
          K. Sakamoto\inst{2}
	 \and
          J. S. Gallagher\inst{3,1}
	\and
        K. Dasyra\inst{4}
       \and
	K. Wada\inst{5}
	\and
         F. Combes\inst{6}
         \and
	  S. Garc\'ia-Burillo\inst{7}
        \and
         L. E. Kristensen\inst{8}
	\and
	 S. Mart\'in\inst{9,10,11}
        \and	
          P. van der Werf\inst{12}
	\and
 	A. S. Evans\inst{13}
	\and
        J. Kotilainen\inst{14}
             }

 \institute{Department of Earth and Space Sciences, Chalmers University of Technology, Onsala Observatory,
              SE-439 92 Onsala, Sweden\\
              \email{saalto@chalmers.se}
\and  Institute of Astronomy and Astrophysics, Academia Sinica, PO Box 23-141, 10617 Taipei, Taiwan 
\and  Department of Astronomy, University of Wisconsin-Madison, 5534 Sterling, 475 North Charter Street, Madison WI 53706, USA
\and Department of Astrophysics, Astronomy \& Mechanics, Faculty of Physics, University of Athens, Panepistimiopolis Zografos 15784, Greece
\and Kagoshima University, Kagoshima 890-0065, Japan
\and Observatoire de Paris, LERMA (CNRS:UMR8112), 61 Av. de l'Observatoire, 75014 Paris, France 
\and Observatorio Astron\'omico Nacional (OAN)-Observatorio de Madrid, Alfonso XII 3, 28014-Madrid, Spain
\and Harvard-Smithsonian Center for Astrophysics, 60 Garden Street, Cambridge, MA 02138, USA
\and European Southern Observatory, Alonso de Córdova 3107, Vitacura, Santiago, Chile
\and Joint ALMA Observatory, Alonso de Córdova 3107, Vitacura, Santiago, Chile
\and Institut de Radio Astronomie Millim\'etrique (IRAM), 300 rue de la Piscine, Domaine Universitaire de Grenoble,
38406 St. Martin d$ ' $H\`eres, France
\and Leiden Observatory, Leiden University, 2300 RA, Leiden, The Netherlands
\and University of Virginia, Charlottesville, VA 22904, USA, NRAO, 520 Edgemont Road, Charlottesville, VA 22903, USA
\and Finnish Centre for Astronomy with ESO (FINCA), University of Turku, V\"ais\"al\"antie 20, FI-21500 Kaarina, Finland
 }

   \date{Received xx; accepted xx}

  \abstract{With high resolution ($0.\asec 25 \times 0.\asec 18$)  ALMA CO 3--2 (345 GHz) observations of the nearby
($D$=21 Mpc, 1\arcsec=102 pc), extremely radio-quiet galaxy NGC1377, we have discovered a high-velocity, very collimated nuclear
outflow which we interpret as a molecular jet with a projected length of $\pm$150 pc.  The launch region is unresolved and lies inside a
radius $r<10$ pc.  Along the jet axis we find strong velocity reversals where the projected velocity swings from -150 \kms\ to +150 \kms. 
A simple model of a molecular jet precessing around an axis close to the plane of the sky can
reproduce the observations. The velocity of the outflowing gas is difficult to constrain due to the velocity reversals but we estimate it to be between
240 and 850 \kms\ and the jet to precess with a period $P$=0.3-1.1 Myr.  The CO emission is clumpy along the jet and the total molecular mass in the
high-velocity ($\pm$(60 to 150 \kms)) gas lies between $2 \times 10^6$ \msun\ (light jet) and  $2 \times 10^7$ \msun\ (massive jet). 
There is also CO emission extending along the minor axis of NGC1377. It holds $>40$\% of the flux in NGC1377 and may be
a slower, wide-angle molecular outflow which is partially entrained by the molecular jet.

We discuss the driving mechanism of the molecular jet and suggest that it is either powered by a (faint) radio jet or by an accretion disk-wind similar to those found
towards protostars. It seems unlikely that a massive jet could have been driven out by the current level of nuclear activity which should then have undergone rapid
quenching.  The light jet would only have expelled 10\% of the inner gas and may facilitate nuclear activity instead of suppressing it.
The nucleus of NGC1377 harbours intense embedded activity and we detect emission from vibrationally excited HCN $J=4-3$ $\nu_2=1f$ which is consistent with hot gas
and dust. We find large columns of H$_2$ in the centre of NGC1377 which may be a sign of a high
rate of recent gas infall.  The dynamical age of the molecular jet is short ($<$1~Myr),  which could imply that it is young and
consistent with the notion that NGC1377 is caught in a transient phase of its evolution.  However, further studies are required to determine the age of the molecular jet,
its mass and the role it is playing in the growth of the nucleus of NGC1377.
 }

    \keywords{galaxies: evolution
--- galaxies: individual: NGC1377
--- galaxies: active
--- galaxies: jets
--- galaxies: ISM
--- ISM: molecules}

 \maketitle

%________________________________________________________________

\section{Introduction}
\label{s:intro}

The growth of central baryonic mass concentrations and their associated supermassive black holes (SMBHs) are key components of galaxy evolution
\citep[e.g.][]{kormendy13}. The underlying processes behind the evolution of the SMBH and how it is linked to its host galaxy and its interstellar gas are,
however, not well understood. In addition, it is not clear how SMBHs can grow despite the energy/luminosity of accretion that leads to gas expulsion from the region. 
Massive molecular outflows powered by AGNs and bursts of star formation are suggested as being capable of driving out a large fraction of the galaxy's cold gas reservoir
in only a few tens of Myr \citep[e.g.][]{nakai87,walter02,feruglio10,sturm11, aalto12a, combes13, bolatto13, cicone14, sakamoto14, garcia14, aalto15a, alatalo15, feruglio15}.
To maintain nuclear activity and growth, an inflow of gas from larger radii is therefore required.

Cold molecular gas has been proposed as an important source of fuel for SMBH growth since the accretion of hot gas is meant to be inefficient and slow \citep{blandford99,nayakshin14}.
However, it is not known how the cold gas is deposited into the inner nucleus of the galaxy.  
This angular momentum problem is similar for the growth of SMBHs and the formation of stars \citep{larson09} and is even more severe for SMBHs because
they are smaller than stars in relation to the size of the system in which they form. Thus, the mass that SMBHs may achieve is likely to be strongly regulated
by the efficiency of angular momentum transfer during the fuel process. In protostars there is strong evidence of a physcal link between infall and outflow
\citep[e.g.][]{arce} and angular momentum can be transferred by molecular jets and outflows. A link betwen infall and outflow seems to also exist for galaxy
nuclei and AGNs \citep[e.g.][]{davies14,garcia14}

Chaotic inflows of cold gas clumps with randomly oriented angular momenta have been suggested
as alternatives to large scale disks in feeding the growth of the SMBH \citep{king07,gaspari13,nayakshin12}. In this scenario, SMBH growth may occur primarily through
multiple small-scale accretion events, rather than continous accretion \citep[e.g.][]{king07} leading to AGN luminosity variations on time scales 
of $10^3 - 10^6$ yr \citep{hickox14}.  A somewhat contrasting picture is that angular momentum may be effectively transported by, for example, bars and spiral density
waves on large and small scales (see e.g. discussion in \citet{garcia14}). AGN luminosity and nuclear growth is therefore expected to vary depending on the interplay between
mode of accretion, outflow, and winds.

To test how gas inflow and the feedback of central activity influences the growth of SMBHs it is important to study galaxies in early or transient phases of their
nuclear evolution. NGC1377 is a likely example of such a system. It belongs to a small subset of galaxies that has a pronounced deviation from the well-known
radio-to-FIR correlation, having excess FIR emission compared to the radio ($q>3$; $q$=log{[FIR/3.75$\times 10^{12}$~Hz]/$S_{\nu}$(1.4GHz)} \citep{helou85}).
These FIR-excess and radio-quiet galaxies are rare. \citet{roussel03} find that they represent a small fraction (1\%) of an infrared flux-limited sample in the local universe,
such as the IRAS Faint Galaxy Sample.  Their scarcity  is likely an effect of the short time spent in the FIR-excess phase, making them ideal targets for studies of transient 
stages of AGN, starburst, and feedback.

\subsection{The extremely radio-quiet FIR-excess galaxy NGC~1377}

NGC~1377 is a member of the Eridanus galaxy group at an estimated distance of 21~Mpc (1\arcsec=102~pc) and has
a far-infrared luminosity of $L_{\rm FIR}=4.3 \times 10^9$ L$_{\sun}$ \citep{roussel03}.
In stellar light, NGC~1377 has the appearance of a regular lenticular galaxy \citep{rc3} although \citet{heisler94} and \citet{roussel06} find
a faint dust lane that extends along the southern part of the minor axis.  

NGC~1377 is  the most radio-quiet, FIR-excess galaxy known to date with radio synchrotron emission being deficient by at least a factor of 37 with 
respect to normal galaxies \citep{roussel03,roussel06}.
Interestingly,  H~II regions are not detected through near-infrared 
hydrogen recombination lines or thermal radio continuum even though faint optical emission lines are present \citep{roussel03,roussel06}.  Deep mid-infrared
silicate absorption features suggest that the nucleus is enshrouded by large masses of dust \citep[e.g.][]{spoon07}.
This supports the notion that NGC1377 may be in a transient phase of its evolution since a more advanced nuclear activity is expected to have cleared out
the enshrouding material.
It has been suggested that the compact IR nucleus may be the site of a nascent ($t<$1~Myr) opaque starburst \citep{roussel03,roussel06}
or of a buried AGN  \citep{imanishi06,imanishi09}. 

High resolution SMA CO 2--1 observations revealed a large central concentration of molecular gas and a massive molecular outflow \citep{aalto12b}
that appeared to be young ($\sim1.4$~Myr).  The extremely high nuclear dust and gas obscuration of NGC1377 aggravates the determination
of the nature of the nuclear activity and the driving force of the molecular outflow, but the extraordinary radio deficiency implies transient nuclear activity .

\medskip
\noindent
We used the Atacama Large Millimeter/submillimeter Array (ALMA) to observe CO 3--2 at high resolution in NGC1377  aiming to
determine the nature of the buried source and the structure and evolutionary status of the outflow.  Here we present the discovery of a high-velocity, extremely collimated and
precessing molecular jet in NGC1377.  Our results show that the nuclear source is likely an AGN and that we are either witnessing a faint
radio jet driving a molecular collimated outflow, or a jet powered by cold accretion.   The nuclear activity of NGC1377 may be fading, or the large
nuclear concentration of gas and dust signify that the major AGN event has not yet occured.  We also discuss how the gas transfer
in the moleular jet may instead foster gas recycling and how this process may promote SMBH growth.

%__________________________________________________________________

\section{Observations}

\label{s:obs}

\begin{figure}[tbh]
\includegraphics[scale=0.77]{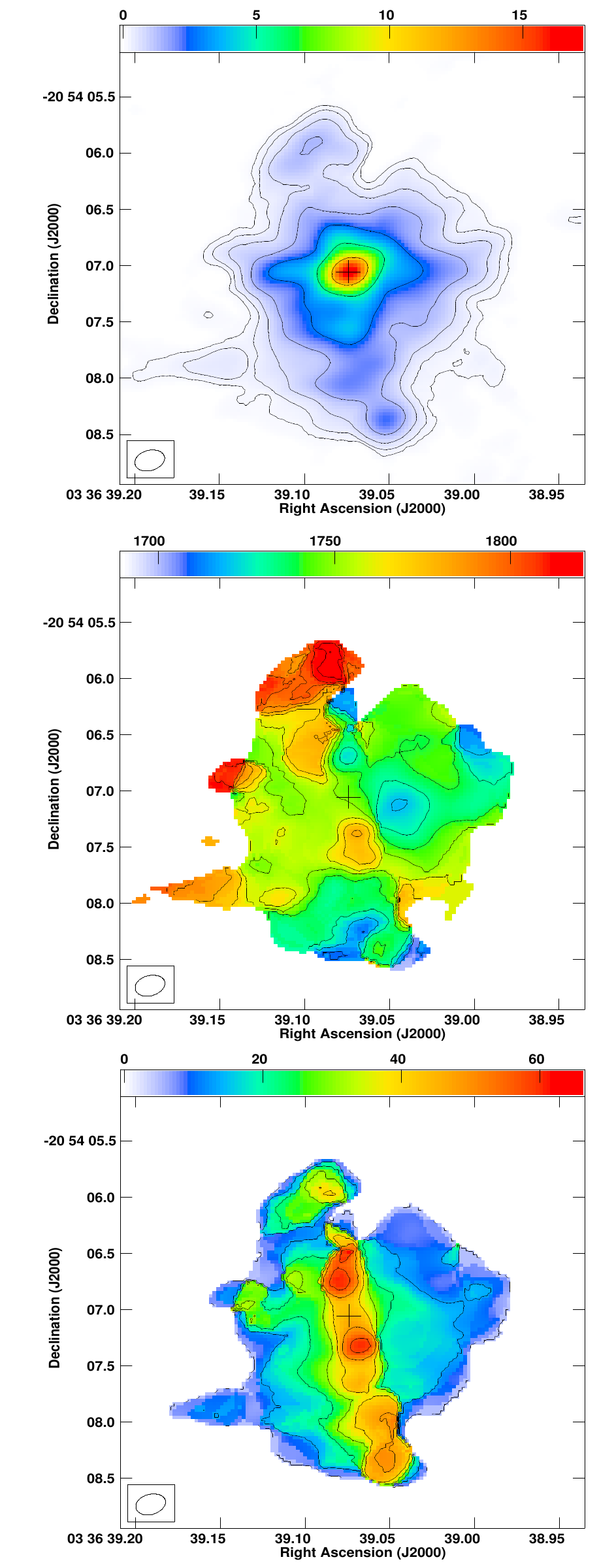}
\caption{\label{f:mom} CO 3-2 moment maps. Left: Integrated intensity (mom0) where contours are 1.7$\times$ (1, 2, 4, 8, 16, 32, 64)  Jy \kms\ beam$^{-1}$. Colours range
from 0 to 172 Jy \kms\ beam$^{-1}$. Centre: velocity field (mom1) where contours range from 
1690 \kms\ to 1820 \kms\ in steps of 10 \kms. Right: Dispersion map (mom2) where contours are 
4.4$\times$(1, 3, 5, 7, 9, 11, 13) \kms. Colours range from 0 to 66 \kms. The cross indicates the position of the 345 GHz continuum peak (see Table~\ref{t:flux}).
}
\end{figure}

\begin{figure}[tbh]
\includegraphics[scale=0.9]{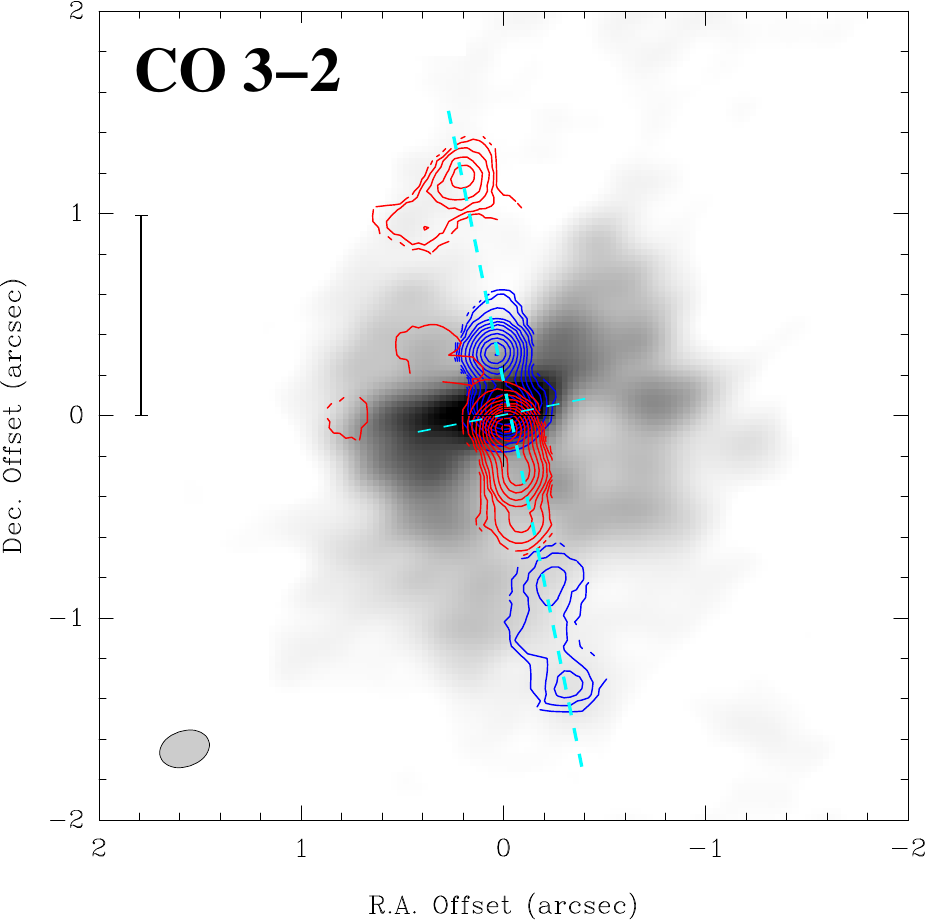}
\caption{\label{f:jet}   CO 3--2 integrated intensity image where emission close to systemic velocity (1700 - 1760 \kms)  is shown in greyscale (ranging from
0 to 70 Jy \kms). The high velocity ($\pm$80 to $\pm$150 \kms) emission from the molecular jet is shown in contours (with the red and blue showing the velocity reversals). 
The contour levels are 1.0$\times$(1, 2, 3, 4, 5, 6, 7, 8, 9)  Jy \kms\ beam$^{-1}$. The dashed lines indicate the 
jet axis and the inferred orientation of the nuclear disk. The CO 3--2 beam is shown as a grey ellipse in the  bottom left corner. The vertical bar indicates a scale of 100 pc.
}
\end{figure}

\begin{figure}[tbh]
\includegraphics[scale=0.48]{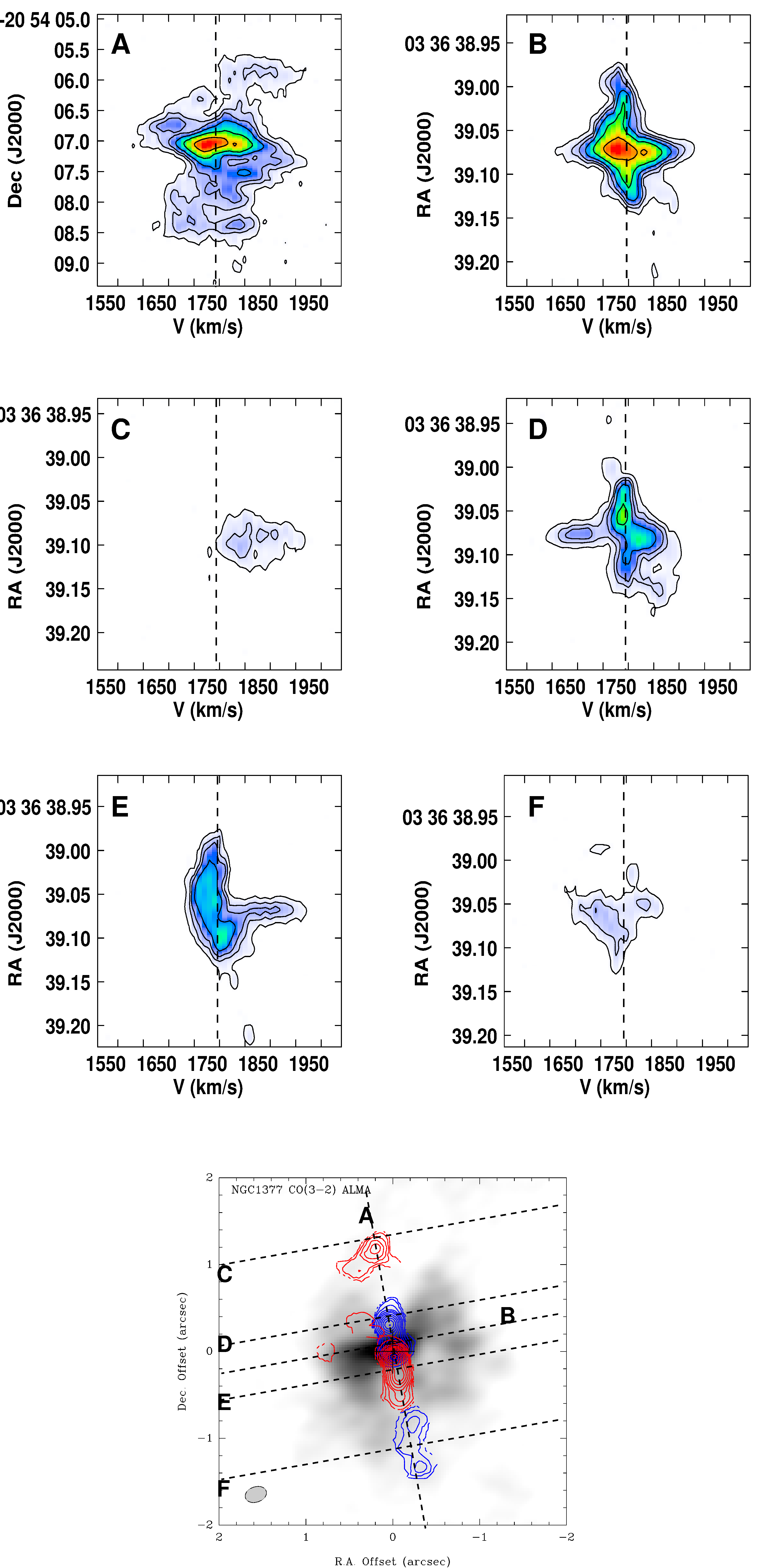}
\caption{\label{f:pv} Position-velocity (PV) diagrams showing gas velocities in five different slits: (A) along the jet axis; (B) Perpendicular to the jet axis through the 
nucleus; (C) Perpendicular to the jet axis at 1.\arcsec 2 to the north; (D) Perpendicular to the jet axis at 0.\arcsec 25 to the north; (E) Perpendicular to the jet axis at 0.\arcsec 25 to the south; (F) Perpendicular to the jet axis at 1.\arcsec 2 to the south.
Contour levels are 3.1$\times$(1, 3, 5, 9, 18, 36) mJy beam$^{-1}$ thus the first level is at 4$\sigma$. The colour scale range from 2 to 156 mJy beam$^{-1}$.
}
\end{figure}

Observations of the CO J=3--2 line were carried out with ALMA (with 35 antennas in the array)
on 2014 August 12, for about half an hour on-source and with good atmospheric conditions
(precipitable amount of water vapour of $\sim$0.5~mm). The phase centre was set to $\alpha$=03:36:39.074 
and $\delta$=$-$20:54:07.055 (J2000).

The correlator was set up to cover two bands of 1.875~GHz in spectral mode, one centred at a
frequency of $\sim$344.0~GHz to cover the CO J=3--2 line (in the lower side band), and the other
centred at 354.3~GHz to cover the HCO$^+$ J=4--3 and HCN $J=4-3$ $v=0$ and $v=1f$ lines (in the
upper side band). The velocity resolution for these bands was 1.0 km/s after Hanning smoothing.
In addition, two 2~GHz bands were set up in continuum mode, i.e., with a coarser velocity
resolution of $\sim$27~\kms, centred at 342.2 and 356~GHz, respectively.

The bandpass of the individual antennas was derived from the quasar $J0423-0120$. The quasar
$J0340-2119$ ($\sim$0.3~Jy) was observed regularly for complex gain calibration.
The absolute flux scale was calibrated using the quasar $J0334-401$.  The flux density
for $J0334-401$ was extracted from the ALMA flux-calibrator database.

After calibration within the CASA reduction package, the visibility set was imported into the AIPS
package for further imaging. The synthesized beam is $0.\asec 25 \times 0.\asec 18$ (25$\times$18 pc for NGC1377)
with Briggs weighting (parameter robust set to 0.5) and the resulting data has a sensitivity of 0.8~mJy per beam in a 10~\kms\
(12~MHz) channel width.

%-----------------------------------------------Results------------------------------------------------------------------------

%---------------------------------------------------------------------------------------------------

\section{Results}

\begin{table}
\caption{\label{t:flux} { CO 3--2 flux densities and molecular masses$^a$}}
\begin{tabular}{ll}
 & \\
\hline
\hline \\ 
Position$^b$ (J2000) & $\alpha$:  03:36:39.073 ($\pm$ 0.\arcsec 01) \\
                           & $\delta$: -20:54:07.05 ($\pm$ 0.\arcsec 01) \\
Peak flux density$^c$ & 156 $\pm$ 1 (mJy\,beam$^{-1}$)\\ 
Flux &  \\
\, \, (central beam) & $17.4 \pm 0.05$ (Jy \kms\ beam$^{-1}$)\\
\, \, (molecular jet)$^d$ & $23.2 \pm 0.5$ (Jy \kms)\\
\, \, (whole map) & $159 \pm 0.5$ (Jy \kms)\\

\hline \\ 
Molecular mass$^e$ & \\
\, \, (central beam) & $1.8 \times 10^7$ \msun\\
\, \, (molecular jet) & $2.3 \times 10^7$ \msun\\
\, \, (whole map) & $16 \times 10^7$ \msun\\
\\

\hline \\
\end{tabular} 

{\it a)}\,  Listed  errors are 1$\sigma$ rms.

{\it b)}\, The position of the peak 345 GHz continuum emission and of the CO 3--2 integrated intensity. The peak
$T_{\rm B}$ is at $\alpha$:03:36:39.072 $\delta$:-20:54:07.06 at $V_c$=1730~\kms. 

{\it c)}\, The Jy to K conversion in the $0.\asec 25 \times 0.\asec 18$ beam
is 1~K=4.6~mJy. The peak $T_{\rm B}$ is 34 K corresponding to 156 mJy.

{\it d)}\, The jet flux is integrated from $\pm$(60 to 200) \kms where the blueshifted flux is 5.5 and the redshifted 17.7 Jy \kms.

{\it e)}\,  The H$_2$ mass $M$(H$_2$)=$1 \times 10^4 \, S({\rm CO} 1-0) \Delta \nu \, D^2$ ($D$ is the distance in Mpc, $S \Delta \nu$ is the
integrated CO 1--0 line flux in Jy \kms) for a conversion factor $N$(H$_2$)/$I$(CO\, 1--0)=$2.5 \times 10^{20}$ $\cmmt$). Since
we have CO 3--2 we have to correct for the frequency dependence of the brightness temperature conversion. If CO 3--2 and 1--0
have the same brightness temperature (thermal excitation, optically thick) the correction factor is 1/9. However, usually the CO
emission is subthermally excited and the brightness temperature ratio is expected to be about 0.5 for a giant molecular cloud. Hence
the correction factor we apply is 1/4.5 and $M$(H$_2$)=$2.2 \times 10^3 \, S({\rm CO}\, 3-2) \Delta \nu \, D^2$. The inferred H$_2$ column density
in the central beam is $N$(H$_2$)=$3 \times 10^{24}$ $\cmmt$.

\end{table}

\subsection{CO 3--2 moment maps}

 The CO 3--2 integrated intensity (moment 0) map, velocity field (moment 1) and dispersion map (moment 2) are presented in Fig~\ref{f:mom}.
We smoothed to two channel resolution, then for the moment 0 map we clipped at the 3$\sigma$ level, and for the moment 1 and
2 maps we clipped at 4$\sigma$. The velocity centroids were determined through a flux-weighted first moment of the spectrum of each pixel, therefore
assigning one velocity to a spectral structure. The dispersion was determined through a flux-weighted second moment of the spectrum of each pixel.
This corresponds to the one dimensional velocity dispersion (i.e. the FWHM line width of the spectrum divided by 2.35 for a Gaussian line profile)

The integrated intensity map shows centrally peaked emission with some structure  extending radially from the centre up to a radius of $\sim$1.\asec 5 (150 pc).
An estimated  11\% of the emission is emerging from the inner 25$\times$18 pc (see Table~\ref{t:flux}). The velocity field is complex and implies that the
maximum velocity shifts occur outside the nucleus. There is evidence for a shallow east-west velocity gradient around the nucleus. The moment 2 map reveals a striking, narrow 3\arcsec\ long feature of high dispersion.

The CO emission clearly delineates two separate structures (Fig.~\ref{f:jet}): an extremely well collimated {\it jet-like} structure, which, essentially, is visible at high velocities
and large-scale emission at low velocities, which surrounds the high velocity jet-like feature. We interpret the high velocity feature as a molecular jet (see Sect. \ref{s:jet})
and we will refer to it as such in the text below.

\subsection{The high velocity gas -- a molecular jet}
\label{s:hivel}

The high velocity (projected velocities 60-150 \kms) gas (Fig.~\ref{f:jet}) is aligned in a $\pm$1.\asec 5 ($\pm$150 pc) long, highly collimated,
jet. It has an unresolved width ($d<$20 pc - set by the limit of our resolution) and a position angle PA=10$^{\circ}$. 
In Fig.~\ref{f:jet}  we show that near the nucleus (within 0.\asec5) emission at redshifted velocities is on the southern side and emission
at blueshifted velocities are found to the north. Further along the axis (beyond 0.\asec 5) this reverses. 
%It is launched in an unresolved region in the nucleus with $r< 10$ pc.

\subsection{Systemic and low-velocity gas}
\label{s:lovel}

The systemic and low-velocity gas (projected velocities 0-60 \kms) consists of a bright central disk-like feature with PA=105$^{\circ} \pm 5^{\circ}$ and  larger scale emission
extending primarily along the minor axis of NGC1377. Along the PA of 105$^{\circ} \pm 5^{\circ}$ there is an east-west velocity shift of $\sim$50 \kms.
The low-velocity emission surrounds the  molecular high-velocity jet in a butterfly-like pattern (Fig.~\ref{f:jet} ).
Most of the CO 3--2 flux of NGC1377 emerges from this minor axis structure (Table~\ref{t:flux}). The minor axis extent of the systemic emission is similar to that of the 
high-velocity molecular jet, but we note that at zero velocities, negatives in the map indicate that some flux is missing from extended emission. The maximum
recoverable scale of our observations is of the order of $\sim5$\arcsec.

\subsection{Position-velocity (PV) diagrams}

In Fig~\ref{f:pv} we present five PV diagrams to show the distinct structure of the high-velocity emission in relation to that of the systemic and low-velocity
gas.

The PV diagram along the jet axis (A) shows the velocity reversals. Near the nucleus the highest velocity is blueshifted to the north and redshifted to the
south. The maximum velocities occur about 0.\asec 25 (25 pc) away from the nucleus. Further away (1.\asec 2) the highest velocity is now redshifted on
the north side and blueshifted to the south.

The PV diagram also shows that the CO emission peaks strongly in the nucleus and that the emission along the jet axis is clumpy.  The clumps are unresolved
in the CO 3--2 beam and from the Jy to K conversion in Table~\ref{t:flux} we find that the clumps have brightness temperatures of $T_{\rm B}$(CO 3--2)=1 -- 8 K. 
The CO 3--2 line widths of the gas clumps in the jet are high ranging from 50 to $\sim$150 \kms, which is evident in the PV diagram along the jet axis, as well as
in the PV diagrams cut across the jet (C-F in Fig~\ref{f:pv}).

We show four PV diagrams oriented perpendicular to the jet: two at distance $\pm$ 0.\asec 25 from the nucleus (D and E) and two 1.\asec 2 from the nucleus (C and F).
They were selected at the locations of highest velocities in the gas along the jet - and also to show the switch in orientation of the high-velocity gas (the velocity reversals).
In PV diagrams D and E the distinction between the narrow, unresolved high-velocity gas from the extended emission (on scales of $\sim 2"$ (200 pc)) of the
low-velocity gas is clear. On the north side the high-velocity blueshifted emission in (D) is narrower than the redshifted high-velocity emission further out (C). A similar
pattern is seen to the south where the narrow redshifted emission near the nucleus (E) is more confined than the blueshifted emission further from the nucleus (F). 
Here, there is also some emission at near-systemic velocities as well as an additional blueshifted component. Comparing D and E we find that the low-velocity gas to
the north is slightly redshifted with respect to systemic velocity and to the south the emission is somewhat blueshifted.

We also present a PV diagram across the nucleus that is perpendicular to the jet component (B). Again, it shows the central concentration of the CO 3--2 emission, broad
unresolved emission on the nucleus, and narrower emission extending to the east and west of the nucleus. There is a low velocity shift from east to west of $\pm$25 \kms.

\subsection{Nuclear gas}
\label{s:nuclear}

\noindent
Velocities in the nucleus span a total of 300 \kms. It is not clear what amount of this constitutes rotation of a circumnuclear disk and what amount
stems from the outflowing gas in the jet.  The velocities in the moment 1 map (Fig~\ref{f:mom}) do not show much rotation around the
nucleus. The nuclear emission is broad but unresolved in space, and the velocity outside of the nucleus drops quickly with radius along the major axis
of the galaxy, as is evident in PV diagram (B) in Fig~\ref{f:pv}.

\noindent
From the CO luminosity, we infer an H$_2$ column density of $N$(H$_2$)=$3 \times 10^{24}$ $\cmmt$ (Table~\ref{t:flux}) towards the nucleus.
This would imply that the nucleus of NGC1377 is Compton thick and similar to the nuclei of other extremely obscured early type disk galaxies, such as
NGC4418 \citep{sakamoto13,costagliola13}, IC860 and Zw049.057 \citep{falstad15,aalto15b}, but more studies are required to confirm this high
$N$(H$_2$) for NGC1377.

\noindent
Apart from CO 3--2 we also detected HCO$^+$ and H$^{13}$CN $J=4-3$ and  vibrationally excited HCN $J=4-3$ $\nu_2=1f$ ($T = E_{\rm l} / k$=1050 K)
The vibrationally excited lines is a factor of 20-30 times fainter than CO 3--2 in the nucleus, but its detection is consistent with a large $N$(H$_2$) and the presence
of very hot gas and dust \citep{aalto15b}.
We also detect lines at redshifted frequency $\nu$=342.26 and 344.5 GHz. The identification of these lines is not clear but we tentatively identify the first as
HC$^{15}$N $J=4-3$ and the second either as vibrationally excited HC$_3$N $J=38-37$ $\nu_4=1$, $\nu_7=1$,  or as SO$_2$. We present spectra and a brief
discussion of the line identification in Appendix~\ref{s:A1}.

\subsection{Continuum}

\begin{figure}[tbh]
\includegraphics[scale=0.3]{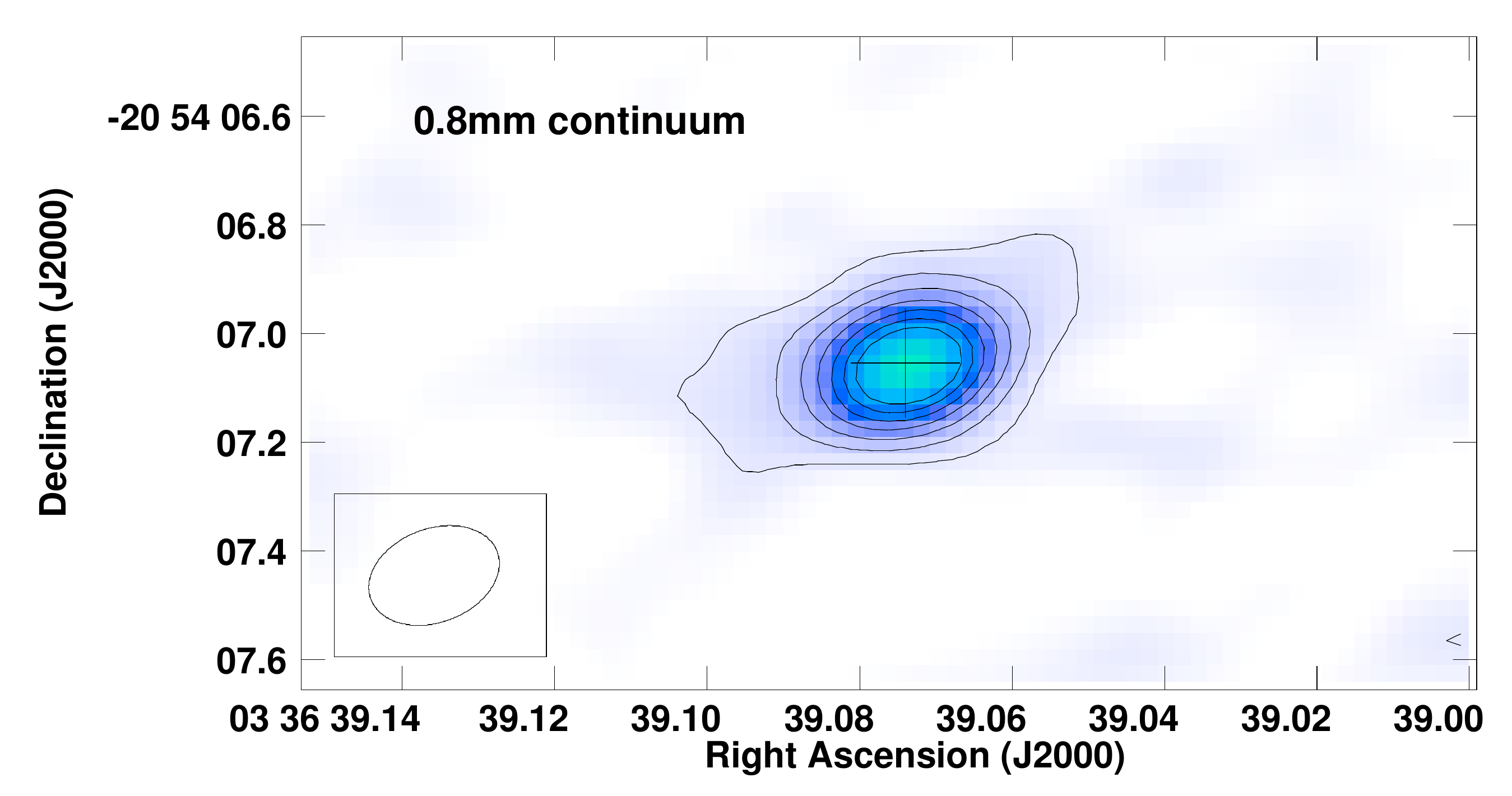}
\caption{\label{f:cont} 0.8mm continuum (merged 342, 349, 356 GHz line-free channels). Contour levels are 0.14 $\times$(1,2,3,4,5,6,7) mJy beam$^{-1}$.  The lowest contour is at 3$\sigma$. The cross indicates the continuum peak position (see Table~\ref{t:flux}).
}
\end{figure}

We merged all line-free channels in our observations into a 0.8mm continuum image (Fig.~\ref{f:cont}). It consists of a compact component and some extended emission.
In the $0.\asec 25 \times 0.\asec 18$ beam, the deconvolved
FWHM size is 0.\asec 25 $\times$ 0.\asec 09 and a position angle PA=104 $\pm$ 5$^{\circ}$. The continuum is faint (1.3$\pm$0.1 mJy beam$^{-1}$ peak and
2.2$\pm$0.3 mJy integrated). The rms is  0.045 mJy. The continuum and CO 3--2 peak in the same position, which we assume is the nucleus of the galaxy.

%------------------------------------- Discussion ----------------------------------------------------------------------------------
%-----------------------------------------------------------------------------------------------------------------------------------

\section{Discussion}

\subsection{The high velocity gas: a precessing molecular jet?}

\label{s:jet}

We interpret the high velocity CO 3--2  emission as emerging from a highly collimated and ordered molecular jet.  
The striking velocity reversals along its symmetry axis are consistent with those of jet precession \citep[e.g.][]{rosen04}.
The maximum velocity swings from 1590  to 1910 north of the nucleus (Fig~\ref{f:pv} figure (A)) and from 1920 to 1650
to the south. Thus on average the shift is 300 \kms. The velocity shifts to the north and south appear fairly symmetric,
which suggests that the symmetry axis of the jet should be relatively close to the plane of the sky and thus launched from a highly inclined disk. 

The 0.8mm continuum image (Fig.~\ref{f:cont}) implies a nuclear disk of  inclination 70$^{\circ} \pm$ 10$^{\circ}$
and a FWHM radius of 13~pc (although we caution that the continuum emission is faint and only marginally resolved).
 In addition, the nuclear CO emission lines are broad with an unresolved dynamics, which is also consistent with the notion of a compact, highly inclined
nuclear disk.

%----

\subsubsection{Simple models}

\begin{figure}[tbh]
\includegraphics[scale=0.35]{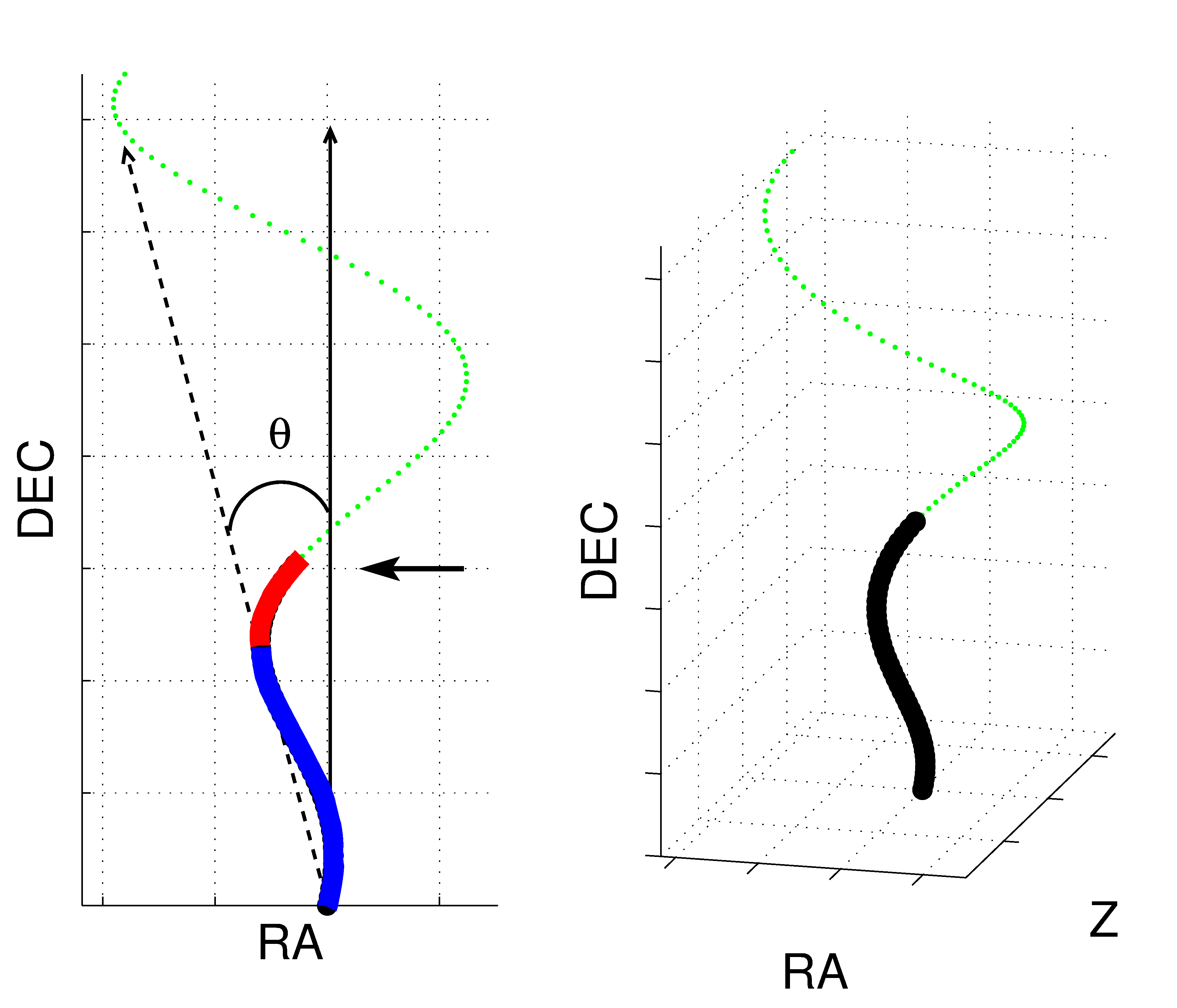}
\includegraphics[scale=0.25]{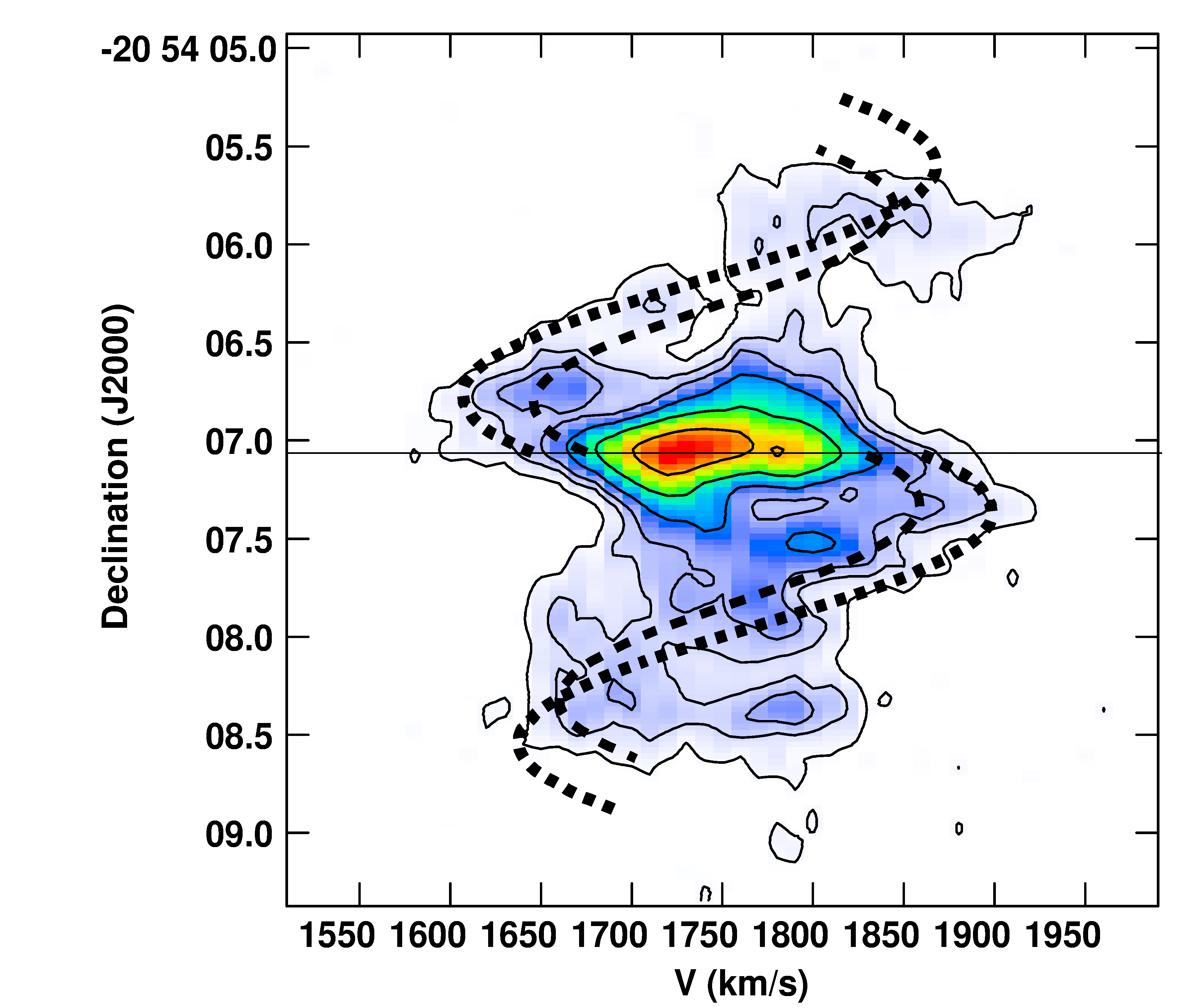}
\caption{\label{f:pvmodel} Simple schematic jet model where we have rotated the jet symmetry axis from PA=10$^{\circ}$ to 0$^{\circ}$. Top: To the right, the northern part
of the jet viewed face on. The curve indicates the pattern of the jet path on the sky and the blue and red colours indicate blue- and redshifted emission in the sight-line. The precession angle here is $\theta$=15$^{\circ}$ and the arrow indicates where the CO 3--2 emission in the jet ends. The right panel shows the jet viewed from
an angle of 45$^{\circ}$ to illustrate its 3D nature. Bottom: The observed PV diagram along the jet axis (Panel (A) in Fig.~\ref{f:pv}) with the superposed tracks of a precessing jet of $\theta$=15$^{\circ}$ and outflowing velocity $v_{\rm out}$=390 \kms\ and $v_{\rm out}$=520 \kms\ indicated with dashed curves. (These values are within the range
for $\theta$ and $v_{\rm out}$ discussed in Sect.~\ref{s:parameters}). We assume $v_{\rm out}$ to be constant and a jet without width. 
}
\end{figure}

\begin{figure}[tbh]
\includegraphics[scale=0.36]{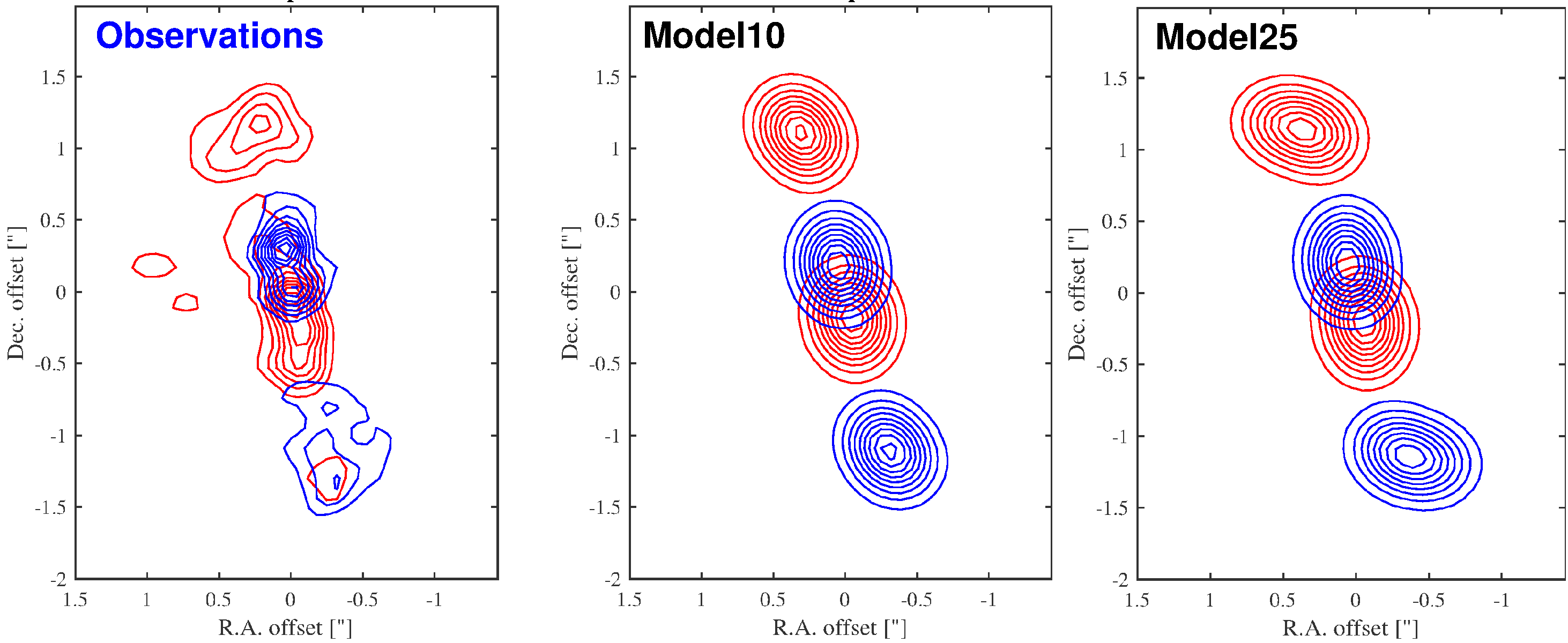}
\caption{\label{f:model}   Contour plots of integrated red- and blueshifted
emission ($>\pm$60 \kms) for ALMA data (right) and model of precessing jet (left): The precession angle is 10$^{\circ}$ (centre) and 25$^{\circ}$ (right) the inclination
of the model precession axis along the line of sight is close to zero.  
}
\end{figure}

The PV diagram along the symmetry axis of a precessing jet of constant outflow velocity shows the projected velocity oscillate\footnote{We use the formalism 
by \citet{wu09}  to describe the line-of-sight velocity change along the symmetry axis of the jet: $v_{\rm LOS}$=$v_{\rm LSR} \pm v_{\rm out}[\cos \theta \sin i + 
\sin \theta \cos i \cos (2\pi l/\lambda +\phi_0)]$, where $v_{\rm LOS}$ is the observed line-of-sight velocity, $v_{\rm out}$ is the outflow velocity in the jet, $i$ is
the inclination of the jet symmetry axis to the plane of the sky, $\theta$=precession angle, $l$=distance from the nucleus, $\lambda$=precession length scale, $\phi_0$=initial phase at the nucleus.} as the jet alternates its direction towards and away from the observer. This is demonstrated schematically  in
Fig.~\ref{f:pvmodel}  where we show the resulting PV diagram of a simple model with a precession angle $\theta$=15$^{\circ}$, the inclination of the precession
axis is zero (i.e. in the plane of the sky), and the outflow velocity $v_{\rm out}$ is constant. The precession has gone through slightly more than half a period.
The first maximum velocity occurs 0.\arcsec 25 above the nucleus when the jet is most pointed towards us, implying that the jet close
to the nucleus is seen at an angle.

We require higher spatial resolution to carry out proper model fits to the jet properties. However, for illustrative purposes we present two model maps in  Fig.~\ref{f:model}, 
showing what the model above would look like if we presented it in contour plot form i.e.  similar to the high velocity contours in Fig.~\ref{f:jet}.
We show two scenarios: one with precession angle $\theta$=25$^{\circ}$ and $v_{\rm out}$=260 \kms, the other with $\theta$=10$^{\circ}$ and $v_{\rm out}$=600 \kms,  to
demonstrate the effect of the precession angle on the high-velocity contour plots.
Here the inclination of the precession axis to the declination axis is 10$^{\circ}$ and, to the plane of the sky, it is zero. We assume a jet width of 0.\arcsec 4.

\subsubsection{Jet parameters}
\label{s:parameters}
The PV diagrams across the jet will show features that are broad in velocity if the cut includes the maximum projected velocity. For these velocities, the spatial
extent will be the lowest and the emission will be narrowest near the base of the jet, while the emission at maximum velocity will be broader
further away from the nucleus owing to the precession of the jet. This effect can be seen in Fig.~\ref{f:pv} (panels (C) and (D)) where
the blueshifted jet component near the nucleus in (D) is narrower than the redshifted jet component further away (to the north) from the
centre. In addition, the position of the redshifted jet component is shifted to the east, compared to the blueshifted component.  This also gives the jet precession
direction implied in Fig.~\ref{f:pv}. The offset of the redshifted jet component to the north can be used to estimate the precession angle. The maximum velocity is
expected to be completely aligned with the jet axis,  but emission at lower redshifted velocities are coming in from the east (also showing the direction of the precession). 
The east-offset implies a precession angle $\theta$=10$^{\circ}$-25$^{\circ}$, but this is, of course, very uncertain since we only have slightly more than half a turn
of the jet. The unprojected outflow velocity $v_{\rm out}$ depends on $\theta$ and the observed maximum projected velocity $v_{\rm proj}$. This may either be
done by selecting the velocity at the 3$\sigma$ contour or the velocity of the brighter clumps. This gives a rough span to $v_{\rm proj}$ of 100-150 \kms.
The outflow velocity should, therefore,  lie in the range $v_{\rm out}$=240-850 \kms\ with a precession period $P$=0.3-1.1 Myr. The dynamical age of the full
length of the molecular jet appears to be short. The jet can be traced out to $\sim$150 pc and for $v_{\rm out}$=240-850 \kms\ the time scale ranges between
$t$=0.2 and 0.7 Myr.
% In the first case, the precession period is $P$=1.3 Myr and, for
%the faster outflow, $P$=0.5 Myr. 

\subsubsection{Launch region}

The molecular jet emerges from the nucleus and its width is unresolved, which results in a launch region of the jet inside $r$=10 pc.
The nuclear rotation is  also unresolved, but from the PV diagram we estimate a rotational velocity of $\sim$110 \kms\ and, if this
occurs ar $r$=10 pc, the rotational timescale is $\sim$1 Myr.  The precession period must be longer than the rotational timescale of
the jet-launching region and hence the jet is very likely launched close to the nucleus, within the inner few pc.

\subsubsection{Origin of precession}

Jet precession may occur in a variety of astrophysical objects, including low-mass star formation in
the Galaxy (L1157 \citet{gueth96,kwon15}), (NGC 1333-IRAS4A \citet{Santangelo15}, (L 1551 IRS 5 \citet{fridlund94}), ( IRAS 16293-2422 \citet{kristensen13});
Galactic micro-quasars  (SS433 \citet{blundell05}  and  1E 1740.7-2942 \citep{luqueescamilla15}), 
and AGN radio jets  \citep[e.g.][]{veilleux93,steffen97,martividal11,pyrzas15}.

Jet precession may be caused by a warped accretion disk \citep[e.g.][]{greenhill03}  i.e. by the misalignment between the spin orientation of the
black hole and the surrounding accretion disk \citep[e.g.][]{bardeen75,lu05} and an accretion flow that is transporting in gas of misaligned angular momentum \citep{krolik15},
but see also the discussion in \citet{nixon13}.  
Alternatively, in a SMBH binary system, jet precession may be caused by geodetic precession of the spin axis of the primary rotating SMBH being
misaligned with the binary total angular momentum, or by inner disk precession (owing to the tidal interaction of an inclined secondary SMBH). 
%Orbital periods for these types of precession range between $10^3 -10^8$ yr and the period should shorten as the two SMBHs grow closer \citep{komossa06,liu07}. 
Interestingly, the presence of a nuclear gas and dust concentration and a precessing molecular jet can aid the coalescence of the SMBHs into resolving
the "final-parsec problem" \citep{milo03,aly15}. The post-starburst spectrum of NGC1377  (Gallagher et al in prep.) could perhaps be linked to a past
merger event that left left an SMBH binary in the heart of NGC1377.  
%The central region of NGC1377 has a post-starburst optical spectrumsuggesting the presence of an
%aging ($\sim$1 Gyr) starburst. It is possible that this declining burst of star formation can be linked to a merger event that has now lost all its dynamical signatures and 

\subsubsection{Other explanations}
In Appendix~\ref{s:A2} we discuss potential alternative explanations for the high-velocity gas emission structure and why we find them less likely (with current information)
than the precessing jet model presented here.

\subsection{Low-velocity gas}
\label{s:low}

The extremely simple jet model cannot explain all the features we see in the PV diagrams (Fig.~\ref{f:pv}). 
Perpendicular to the jet axis (panels C-F in Fig.~\ref{f:pv}), we see the jet emission as a broad velocity feature and narrow in space. However, there is
also more spatially extended emission at low velocities (panels (D) and (E)).  In PV diagram (A), the lower velocity emission occurs
as straight lines to the north (in particular) but also in the south. There is also an extra component at 1800 \kms\ to the south next to the blueshifted part of the jet.
This emission cannot be directly explained by a simple model of a precessing jet and may emerge from a background disk, a molecular wide-angle wind, or it is
caused by interaction and entrainment by the jet. 

For example a  bow shock can arise by the formation of an internal working surface within the jet at positions of strong velocity
discontinuity, and as the high velocity jet interacts with the surrounding medium \citep{raga93,gueth99, cliffe96,santiago09}. 
%Models by \citet{cliffe96} show that a precessing jet may have global bow shocks wider than those of straight jets. The bow shock can envelope th
%whole structure and may be quite wide (the "shroud"). 
The structure and velocity of the ambient gas may become complex owing to, for example, the action of
the global bow shock and gas sweeping into the wakes of the jet turns. Dynamical simulations of precessing gaseous jets have been carried out by
\citet{raga01}. They present PV diagrams perpendicular to their simulated jet (their Fig. 5) and  find that the transverse spatial extent of the emitting region
is larger at lower radial velocities. In their simulations, this is due to  the presence of bow shock wings trailing behind each internal working surface.
These bow shocks result in transverse extended emission of low radial velocities which forms a 'halo'  component. 
There is a striking similarity between the Fig. 5 of \citet{raga01} and our PV diagrams that are perpendicular to the jet.

The low-velocity gas has redshifted velocities north-east of the jet and blueshifted velocities to the south-west. The angle between the
most red- and blueshifted gas is PA=40-45$^{\circ}$ and the velocity shift is $60 \pm 20$ \kms. Apart from this gradient, there is no significant net shift in velocity
between the north and the south (with deviations at the ends of the jet and at the edges of
the map). There is a small (10-20 \kms) east-west gradient which is somewhat larger (50 \kms) at the disk major axis. 

The PA=40-45$^{\circ}$ velocity structure can be caused by the jet entraining and accelerating a very slow, wide-angle minor axis molecular outflow and/or
that it is interacting with gas already  entrained before. Another possibility is that there is a wind, which is unrelated to the jet and which originates in a disk warped about
20$^{\circ}$, compared to the nuclear disk. This orientation is however not consistent with that of the optical dust absorption features south of the nucleus of NGC1377
( Fig.~1 in \citet{roussel06}, Fig.4c in \citet{heisler94}). The dust structures have a v-shaped morphology (opening angle of $\sim$90$^{\circ}$) and are oriented
almost perpendicular to the stellar disk. They may be caused, forexample, by the precessing jet bow shocks.

\subsection{Comparing previous results for NGC1377}

In our previous paper on NGC1377\citep{aalto12b}, we suggested that the molecular outflow seen in CO 2--1 is biconic with an opening angle of
60$^{\circ}$- 70$^{\circ}$, an outflow mass $>1 \times 10^7$ \msun, and an outflow velocity of 140 \kms. These observations were carried out with three times poorer spatial
resolution and about ten times lower flux sensitivity than the ALMA CO 3--2 data presented here.  The CO 2--1 dispersion map has a cross-like structure that
we used as a basis to suggest the biconic outflow. In the ALMA data, high dispersion is found only along a structure that we now identify as a molecular jet.

It is interesting to note that the position angle of the outflow in the lower resolution CO 2--1 map is around PA=40$^{\circ}$, while the CO 3--2 jet has a
PA of 10$^{\circ}$.  The lower resolution SMA data has likely picked up the velocity shift in the low-velocity gas discussed above (Sect.~\ref{s:low}) which
we propose is caused by jet entrainment (or, less likely, an inclined wide-angle flow). Further studies will reveal more on the origin of this gas component.

%-------------------------------------------------------------------------------

\subsection{Mass and outflow rate}
\noindent
{\it The molecular jet:} \, The molecular mass in the high velocity gas is estimated as $M_{\rm j}({\rm H}_2)=2.3 \times 10^7$ \msun, assuming a standard
CO to H$_2$ conversion factor (see Table~\ref{t:flux}).
For $v_{\rm out}$ between 240 and 850 \kms,  we estimate the mass outflow rate in the jet at  9 -- 40 \msun\ yr$^{-1}$.
This results in a momentum flux of (14 - 200)$L/c$, which is very high and exceeds values typically seen in cases of AGN feedback \citep{cicone14, garcia14}.
However, since we use a standard conversion factor, the H$_2$ mass may have been overestimated. If the gas is turbulent, and the individual gas clouds unbound, the conversion
factor may have to be adjusted down by a factor of 10 \citep[e.g.][]{dahmen98}. 
%The line widths are indeed broad in the molecular jet of NGC1377 (Sec.~\ref{s:hivel}),
%but this does not necessarily reflect the internal velocity dispersion of individual clumps in the jet, but instead an effect of precession/projection/resolution. 
%The  imprint of the disk rotation will also be carried out with the jet.

\smallskip
\noindent
{\it The low-velocity outflow:} \, The mass and velocity in the low-velocity outflow (Sect.~\ref{s:low}) is difficult to estimate since there is the possibility of contamination by
a background disk and the morphology and velocity structure are complicated. But if we assume that all the CO 3--2 flux above and below the stellar disk belong to the
slow outflow, it would constitute 40\% of the total CO 3--2 flux detected in NGC1377. For a standard conversion factor (Table~\ref{t:flux}) this implies
$M_{\rm slow}({\rm H}_2)=6 \times 10^7$ \msun. About a third of this is associated with the entrained (alt inclined wind) part of the flow, with projected velocities 
$\pm 30$ \kms. A generous estimate of $v_{\rm out} \sim$50 \kms\ over 100 pc  implies that 10 \msun\ yr$^{-1}$ may be lifted off the midplane of NGC1377. This number is
highly uncertain.

%---------------------------------------------------------------------------------

\subsection{What is powering the molecular jet?}

\subsubsection{Accretion}
\label{s:accretion}

Jets are generally identified with accretion \citep{blandford98, konigl00,hujeirat03,sbarrato14} and are likely launched by magnetohydrodynamic processes
from the accretion disk and/or the central object. The molecular mass is a crucial ingredient in determining the energetics, nature, and evolutionary stage of the molecular jet. 
We have to resort to a CO to $M$(H$_2$) conversion factor to determine the molecular mass and we have two limiting cases: A {\it massive jet} where 
$M_{\rm j}$(H$_2$) $\sim 10^7$ \msun\ or a {\it light jet} with
$M_{\rm j}$(H$_2$) $\sim 10^6$ \msun. Below we discuss possible driving scenarios in relation to a massive or a light jet. 

%For the molecular jet of NGC1377 we discuss two main scenarios: A. entrainment by a radio jet and B. cold accretion.

\medskip
\noindent
{\it Entrainment by a radio jet } \, Powerful radio jets are launched when an SMBH is growing through hot accretion which is an inefficient accretion
at low rates ($<$1\% Eddington) \citep{mcalpine15}.  This is also referred to as radio mode AGN feedback. Radio jet production has been found for
high Eddington rates where the jet powers do not exceed the bolometric luminosity of their AGNs \citep{sikora13}. The jet may entrain molecular gas from
the disk of the host galaxy (NGC1266 \citep{alatalo11},   IC5063 \citep{morganti15}, M51 \citep{matsushita07} and NGC1068 \citep{garcia14}) or the
molecular gas may form in the jet itself through rapid post-shock cooling \citep{morganti15}.
Observed molecular gas distributions associated with these jets tend to be patchier than the more coherent molecular structure of NGC1377. 
A relativistic radio jet ploughing through a thick disk of gas, is likely to heat and ionise it, and thus form a wide cocoon of multi-phase
and turbulent gas mixture, as simulated by \citet{wagner11}). As shown in \citet{dasyra15}, this kind of cocoon is both pushing on the surrounding gas and
has forward and scattered flows that may lead to complicated velocity patterns. 
%Although it is not clear how this model could explain the highly
%ordered jet motions in NGC1377,
%including a jet-ISM interaction in the form of a hot cocoon is likely important for a full understanding of the gas motions in the radio-jet scenario

\smallskip
\noindent
However, NGC1377 is the most radio-quiet (with respect to the IR luminosity) galaxy found so far and its radio power is very low. (A similar
case with faint radio emission associated with molecular jets may be the double, collimated bipolar outflows of the luminous merger
NGC3256 \citep{sakamoto14}). 
We can use the limit to the 1.4 GHz radio luminosity \citep{roussel06} and the relation between jet power and 1.4 GHz luminosity \citep{birzan08} 
to estimate the energy in a potential radio jet in NGC1377.  We find that it amounts to $<$10\% of the mechanical energy in the massive molecular jet.
A short burst of hot accretion in the nucleus may have led to the formation of a radio jet that then faded very rapidly without re-acceleration of
electrons in the jet itself. If the synchrotron life time  is $t_s=8 \times 10^8\, B^{-2} \gamma^{-1}$  (where $B$=$B$-field, $\gamma$=Lorentz factor, \citep{xu00})
a reasonable combination of $B$ and $\gamma$ can result in a jet lifetime of 0.5-1 Myr.  Also, it is conceivable that heavily mass-loading a radio jet with dense molecular gas
may lead to the quenching of the non-thermal radio emission.  In addition, \citet{godfrey16} recently suggested that jet power and radio luminosity may only be weakly correlated
for cases where the jet energy is being used to, for example, drive shocks.  

In the case of the light jet it is feasible that there would be enough radio power to carry the gas out without invoking a fading or underluminous radio jet.

%------------

\medskip
\noindent
{\it Cold gas accretion} \, The jet may be a hydromagnetic disk-wind (or an accretion X-wind) similar to the extremely collimated molecular outflows found in accreting
low-mass protostars \citep[e.g.][]{konigl00,codella14, kristensen15}. Its torque could efficiently extract disk angular momentum and gravitational potential
energy from the molecular gas. The jet may be powered by accretion onto the central object and/or infalling gas onto the nuclear disk. 

\smallskip
\noindent
Assuming that the $5 \times 10^9$ \lsun\ of NGC1377 emerges from a growing SMBH, the accretion rate would be $\sim 10^{-3} -10^{-2}$ \msun\ yr$^{-1}$  ($L$=${\epsilon \over c^2} {dM \over dt}$ where $\epsilon$=0.1 onto a $10^6$ \msun\ SMBH). This is 10\% of the inferred Eddington luminosity of the SMBH \citep{aalto12b} and is a relatively high rate, placing it in the quasar mode of accretion \citep{mcalpine15}.  But it may require an Eddington or super-Eddington accretion rate to produce the mass-outflow rate we see (even in the case of the light jet), implying that the level of SMBH accretion has dropped recently.

A jet may also be powered by accretion onto a nuclear disk. 
The wind energy is derived from the gravitational energy released from the disk through gas rotation and a coupled magnetic field.
The extracted angular momentum allows cold molecular gas to sink further towards the nucleus. 
In the case of the massive jet, it is not clear how the current rotational energy of the disk could continue to sustain
the outflow since $M_{\rm j}$  would be equal to that in the disk inside its launching region. The binding energy of the jet is similar to the
binding energy of the disk and the outflow speed is at least twice that of the rotational velocity (unless the jet is actually launched very close to
the nucleus from a Keplerian disk).  In the case of the light jet, however, the jet-binding energy  would be much less than that of the disk, and there
would be enough rotational energy to sustain the outflow.

\smallskip
\noindent
We note that the molecular jet is observed as being lumpy which may be due to internal and external shocks, or the condensations are gas clumps that originate in separate accretion/outflow
events. If so, the energetics of the outflow may be different to that of the steady flow scenario we assume above.

%----------------

\subsubsection{Other scenarios}

\noindent
{\it Radiation pressure from dust?}  \, 
Recent work by \citet{ishibashi15} suggests that large momentum flux
outflows ($>10 L/c$) can be obtained in radiation pressure driven outflows if radiation trapping is taken into account. However, it is not clear how radiation pressure
would result in a jet-like feature since it should give rise to a more wide-angle wind.  \citet{wada15} finds that dusty, biconical outflows (opening angles 45$^{\circ}$ -- 60$^{\circ}$)
can be formed as a result of the radiation feedback from AGNs. It is conceivable that this may be happening in NGC1377, in addition to the jet.

\smallskip
\noindent
{\it Starburst winds?} \, In \citet{aalto12b} we discuss the faintness of the star formation tracers (such as optical, NIR and radio emission) of NGC1377. We find that the upper limits on,
for example, the 1.4 GHz continuum imply that star formation falls short by at least one order of magnitude in explaining the momentum flux in the molecular outflow detected with the SMA. 
%Radio continuum emission has remained undetected in NGC1377 until recently when 1.5 and 9 GHz emission was found with the
%Jansky Very Large Array (JVLA) (Costagliola et al. in prep.). The emission is very faint and its spectral index suggests synchrotron- rather than thermal free-free emission
%from HII regions. With the new 1.4 GHz flux we find that supernovae emission is too faint by a factor of 50 to drive the outflow. 
%We find no evidence of an embedded nascent pre-supernova starburst. 
%In addition, the high degree of collimation and strong precession of the molecular jet also imply that it is not powered by the winds of an embedded stellar cluster.

\subsection{Is the molecular jet signaling nuclear growth or quenching?}
\label{s:growth}

There is large molecular mass in the nucleus of NGC1377, which appears to be linked to a current SMBH accretion at a respectable rate of  $\sim$10\% Eddington. So
the question is: has the molecular jet action quenched the nuclear activity, or did it promote it? 

\smallskip
\noindent
{\it Light jet:}\,  Both scenarios discussed in Sect.~\ref{s:accretion}  could power the jet and enable SMBH accretion. A light jet has removed only 10\% of the disk mass while 
it may have transported a substantial amount of angular momentum away from the gas in the disk, allowing it to sink closer to the SMBH.  The molecular jet offers a way for the
cold gas to shed itself of excess angular momentum, which could promote nuclear accretion from a disk. In this scenario, the inflowing gas clouds do not have to have randomly
oriented angular momenta to facilitate accretion.  There is no evidence that star formation has hindered the gas flow toward the nucleus of NGC1377.  Instead there appears to be a mechanism that prevents stars from forming in the high gas surface density nuclear region. Higher resolution studies will hopefully find and resolve the inflowing gas component  in NGC1377.

%\citet{nayakshin12} suggest that SMBHs grow through the chaotic accretion of gas clouds with small and randomly distributed angular momentum. 
%They suggest that a planar mode of %feeding may not be an important mode of SMBH growth since such disks are unstable to gravitational fragmentation. 
%In addition, there appears to be a mechanism that prevents stars from forming
%even in high gas surface density disks since the nucleus of NGC1377 has a very low star formation rate that has yet to reach the SMBH.

\smallskip
\noindent
{\it Massive jet:}\, Current rates of accretion would be difficult to reconcile with a large mass outflow rate. Nuclear activity in the form of radio luminosity, or other forms
of accretion luminosities, are low and are, perhaps, a signature of quenching.  The turning off of the nuclear activity would have to have been abrupt since the molecular
jet can be traced almost all the way down to the centre. Furthermore,  the large masses of molecular gas are surprising since it is not clear why the activity would turn off with
30\% of the nuclear fuel still in place. A possible explanation could be that there has been a recent substantial inflow of molecular gas.

\smallskip
\noindent
The discussion above rests on the assumption that most of the FIR emission originates near the SMBH and is the result of the accretion. However, if the FIR emission is,
instead, related to the jet-ISM interaction in an extremely dense medium, then the SMBH would be in the hot accretion mode instead, but with its synchrotron quenched by the
interaction. If so, we are witnessing the early stages of jet feedback before it has cleared its environment.

\subsubsection{What is the fate of the molecular gas?}

A precessing jet has the potential to impact and stir up a large volume of ambient gas. In NGC1377 the jet appears to entrain gas in a slow moving
outflow, possibly in combination with a wide-angle wind. It is, however, unlikely that the gas in the low-velocity outflow can leave NGC1377 since even an
optimistic estimate of its outflow speed is below the bulge escape velocity $v_{\rm esc}$ for NGC1377 \citep{aalto12b}. Instead, gas may circulate back to the midplane of
NGC1377 where it could eventually participate in star formation or another cycle of nuclear growth. 
%Even if only 10\% of the total outflowing gas returns, and ends up accreting onto the SMBH, the black hole mass will be above that expected from the $M - \sigma$ relation. 

The molecular jet appears to be a young structure with a dynamical age $<$1~Myr (Sect.~\ref{s:parameters}). The estimated $v_{\rm out}$=240-850 \kms\  
is higher than $v_{\rm esc}$ for NGC1377. We find no high-velocity molecular gas outside 200 pc and this would be consistent with the notion that the jet
has been caused by a recent accretion event in the nucleus. This would also be consistent with the high nuclear concentration of molecular gas. However, if the molecular gas
becomes dissociated at 200 pc, we may simply be observing the inner denser part of an older outflow event and, if the gas is not slowing down, it may escape the galaxy.
Yet another alternative is that the jet is rapidly decelerating and its gas is grinding to a halt at its end.  The v-shaped optical dust lane is roughly 2-3 times longer than the
molecular jet/outflow structure, which implies that the molecular jet is part of a somewhat older structure.

\smallskip
\noindent
Our results demonstrate that outflows/jets even from low-power AGNs can have substantial impact on the evolution
of the galaxy, also beyond the innermost pc. We require the high resolution, dynamic range and sensitivity of ALMA to reveal the presence of the molecular jet and
to separate it from surrounding emission. Determining the molecular mass in the jet will provide an important clue as to whether the jet is a signature of growth or
quenching of the nuclear activity. More detailed studies will also reveal how the jet impacts its environment and entrains gas and dust.

\section{Conclusions}

With high resolution (0.\asec2 $\times$ 0.\asec 18)  ALMA CO 3--2 observations of the nearby extremely
radio-quiet galaxy NGC1377, we have discovered a high velocity, collimated molecular jet with a projected length of $\pm$150 pc. 
Along the jet axis we find strong velocity reversals where the projected velocity swings from -150 \kms\ to +150 \kms. 
A simple model of a molecular jet precessing around an axis close to the plane of the sky can
reproduce the observations. The velocity of the outflowing gas is difficult to constrain due to the velocity reversals but we estimate it to be between
240 and 850 \kms\ and the jet to precess with a period $P$=0.3-1.1 Myr.

The jet is launched close to the nucleus inside a radius  $r<10$ pc and its molecular mass lies between
$2 \times 10^6$ ({\it light jet}) and $2 \times 10^7$ \msun\ ({\it massive jet}) depending on which CO to $M$(H$_2$) conversion factor is adopted.
There is also a wide-angle structure of CO emission along the minor axis which may be a slower molecular outflow. 
A substantial fraction of the CO flux is located here and the estimated mass of the minor axis outflow is $6 \times 10^7$ \msun. Its velocity structure is
consistent with parts of the wind being entrained by the jet, or that there is a molecular wind inclined by 30$^{\circ}$ with respect to the jet.

We discuss potential powering mechanisms for the molecular jet. It may be gas entrained by a very faint radio jet, or it is driven by an
accretion disk-wind similar to those found in protostars.  It is important to better constrain the jet molecular mass. Given the possibility of either a
light or a heavy jet, it is difficult to draw conclusions on whether the jet is quenching the nuclear activity or, instead, is enabling it. 
The nucleus of NGC1377 harbours intense embedded activity and, if the current IR luminosity is powered by a growing SMBH, it would have
an accretion rate of $\sim$10\% Eddington. But the origin of the FIR luminosity still needs to be determined.
The light jet would only have driven out 10\% of the nuclear gas which should not (yet) significantly impact the fueling of the activity. It seems, however,
unlikely that a massive jet could have been powered by the current activity and this may be a sign of rapid quenching. 
In this case, the large mass of H$_2$ in the nucleus is surprising and may be caused by a recent massive influx of gas. A fraction of the outflowing gas
may return to the inner region of NGC1377 to fuel further nuclear growth.

NGC1377 is the first galaxy with evidence for a precessing, highly collimated molecular jet. The extreme $q$-value for NGC1377,  the short
apparent time-scale of the molecular jet ($<$1~Myr), and the gas-rich nucleus are all signs consistent with the notion that we are seeing NGC1377 in a transient
phase of its evolution.

NGC1377 offers a unique opportunity for detailed studies of the processes that feed, promote and
quench nuclear activity in galaxies. Further studies are required to determine the age of the molecular jet, driving mechanism, 
its mass and the role it plays in the growth of the nucleus of NGC1377.

\begin{acknowledgements}
This paper makes use of the following ALMA data:
    ADS/JAO.ALMA\#2012.1.00900.S. ALMA is a partnership of ESO (representing
    its Member States), NSF (USA) and NINS (Japan), together with NRC
    (Canada) and NSC and ASIAA (Taiwan), in cooperation with the Republic of
    Chile. The Joint ALMA Observatory is operated by ESO, AUI/NRAO and
NAOJ.
We thank the Nordic ALMA ARC node for excellent support. SA acknowledges
support from the Swedish National Science Council grant 621-2011-4143. 
F.C. acknowledges support from Swedish National Science Council
grant 637-2013-7261 
KS was supported by grant MOST 102-2119-M-001-011-MY3
SGB thanks support from Spanish grant AYA2012-32295. JSG thanks the Chalmers University
for the appointment of {\it Jubileumsprofessor} for 2015. SA thanks S. K\"onig, {\AA}. Hjalmarson,
R. Haas and S. Bourke for discussions of the manuscript.

\end{acknowledgements}

\bibliographystyle{aa}
\bibliography{n1377_ALMA_aalto}

\begin{appendix}

\section{Spectra}
\label{s:A1}

\begin{figure}[tbh]
\includegraphics[scale=0.50]{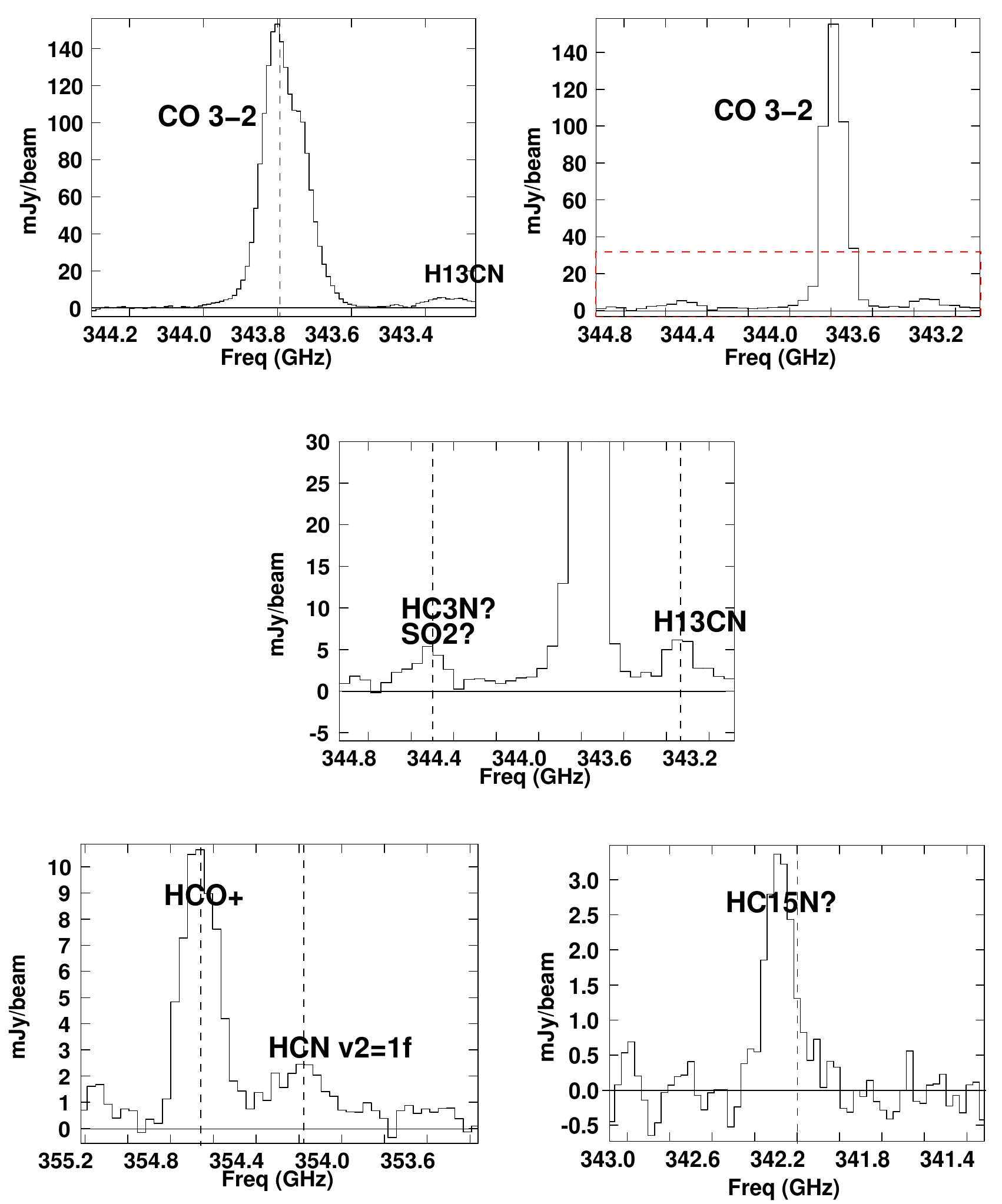}
\caption{\label{f:spec}  Spectra of the nuclear emission in NGC1377. Dashed vertical line indicates $v$=1740 \kms. Top panels: CO 3-2 in high- (left) and low- (right)
resolution spectral mode.  The dashed red box of the right panel indicates the zoomed-in region in the next panel. Centre:  Zoomed-in spectrum showing detections of H$^{13}$CN $J=4-3$ and either HC$_3$N 
$J=38-37$ $\nu_4=1$, $\nu_7=1$ or SO$_2$ 16(4,12)-16(3,13). Bottom: Panels showing detections of HCO$^+$ $J=4-3$, HCN $J=4-3$ $\nu_2=1f$ (left) and a line that we tentatively identify as HC$^{15}$N $J=4-3$. All spectra apart from in panels 2 and 3 have Gaussian smoothed with FWHM of two channels. In panels 2 and 3
there has been no smoothing but frequency resolution is reduced by a factor of 3 (these data stem from another spectral window than that presented in the
first panel).
}
\end{figure}

In Fig.~\ref{f:spec}  we present spectra towards the nucleus of NGC1377. Apart from CO 3--2 we detect HCO$^+$, H$^{13}$CN $J=4-3$,  
vibrationally excited HCN $J=4-3$ $\nu_2=1f$ ($T = E_{\rm l} / k$=1050 K). We detect a line at $\nu$=345.5 GHz which is either  vibrationally excited HC$_3$N $J=38-37$ $\nu_4=1$, $\nu_7=1f$ ($T = E_{\rm l} / k$=1891 K) or it is SO$_2$ 6(4,12)-16(3,13) ($T = E_{\rm l} / k$=148 K).  In addition we detect a line at redshifted frequency $\nu$=342.26 GHz which we tentatively identify as HC$^{15}$N $J=4-3$.  In this case, the line would peak at $v$=1670 \kms\ and thus be blueshifted with respect to the other lines by
60 \kms. This type of shift could be caused by excitation, optical depth and/or abundance gradients and should be investigated in further studies since it may
hold another clue to the nature of the nuclear emission of NGC1377.

\section{Other potential explanations to the high velocity CO 3--2 emission}
\label{s:A2}

\noindent
{\it An orbiting object and/or two jets?} \, Velocity variations in a PV diagram may also be caused by a jet launched from an orbiting object. In this case the velocity
reversals can be dominated by the orbital motion  in a near edge-on plane of rotation.   A possibility would be a jet launched from one of two orbiting
SMBHs. \citet{masciadri02} have discussed the similarities and differences between orbiting and precessing jets.  However, without jet precession the velocity pattern will not fit the structure we see in the observed PV diagram - unless the jet symmetry axis is misaligned with respect to the axis of the plane of rotation. 
Both SMBHs could have jets and a combination of orientation and length of the jets could be put together to reproduce the observed PV diagram. However,
this seems unlikely compared to the relatively simple scenario of one single precessing jet. 

\medskip
\noindent
{\it Jet shocks?} \, A pulsed jet will have a sawtooth like pattern in its PV diagram along the jet major axis \citep[e.g.][]{santiago09}. This pattern is caused by axial compression
and lateral ejection of material inside the internal working surface.  \citet{santiago09} point out that
the effects are localized within the jet so it is not obvious how it would give rise to the large scale shifts and gradients seen here. However, internal and external shocks
would be important for the jet of NGC1377 and thus influence its velocity structure. 

\medskip
\noindent
{\it A bicone projection?} \, Is it possible that the velocity reversals we observe in the PV diagram (A) (Fig~\ref{f:pv}) is a projection effect, instead of the emission from a  collimated jet?  A tilted wide-angle  biconical outflow may result in projected,  foreshortened blueshifted emission from the lower end of the cone, and redshifted (more elongated) emission from the back side of the cone (and vice versa on the underside of the cone). PV diagrams of these scenarios are, for example, presented by (e.g.) \citet{cabrit86}, \citet{das05} and \citet{storchi10}. 

\begin{figure}[tbh]
\includegraphics[scale=0.44]{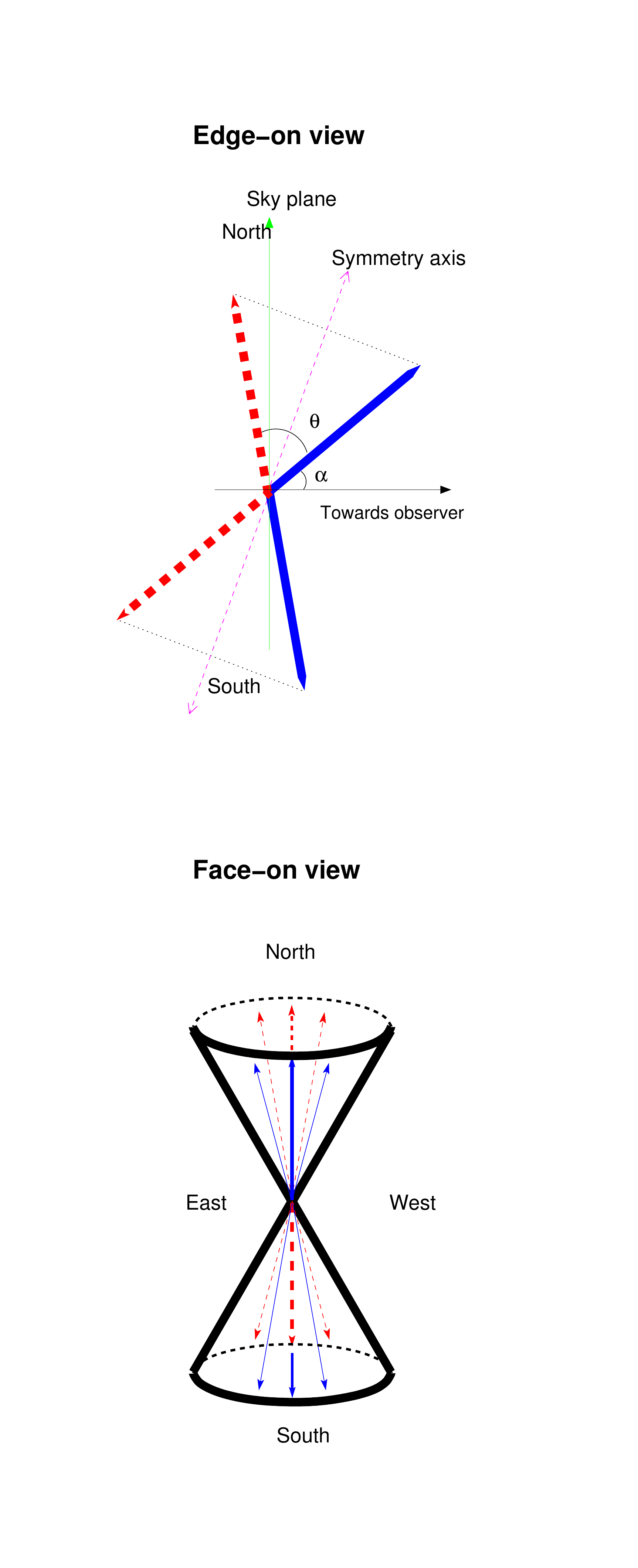}
\caption{\label{f:cone}   Sketch of a hollow cone with opening angle 60$^{\circ}$ and tilt angle 40$^{\circ}$ towards the observer.
}
\end{figure}

\begin{figure}[tbh]
\includegraphics[scale=0.44]{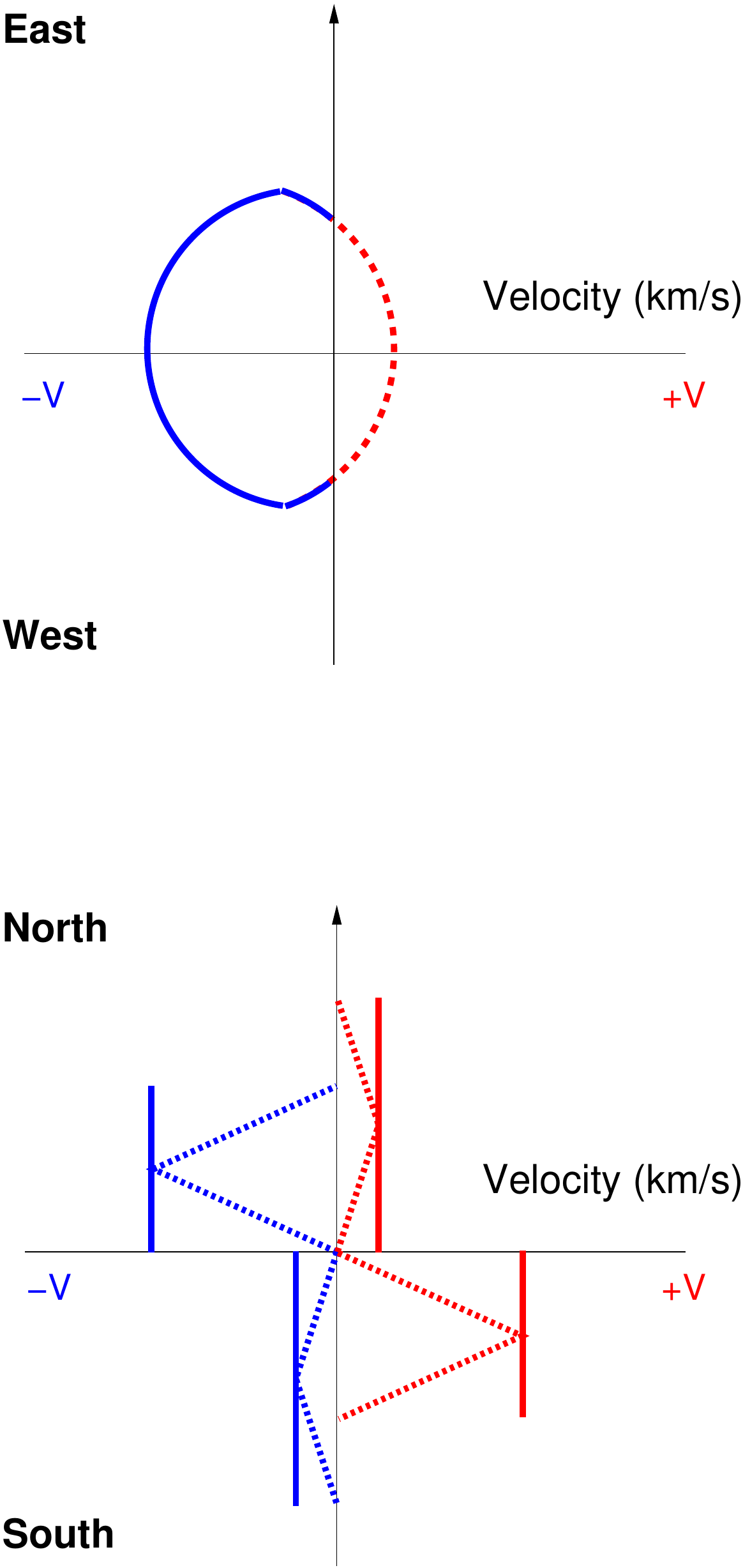}
\caption{\label{f:cone_pv}   Schematic PV diagrams of a cone with its top (northern) part tilted towards us. Top panel: Cut perpendicular to the cone axis showing
the elliptical PV diagram through the cone. Bottom panel: Cut along the cone major axis. We show two simplified cases: The straight solid lines show the PV diagram
of outflowing gas along the cone walls of constant velocity. The dashed lines show the generic PV diagram along the axis of a cone where the
gas is first linearly accelerating and then decelerating to zero velocity.
}
\end{figure}

In Fig.~\ref{f:cone} we show a sketch of a cone (displayed from two angles) with the northern part tilted towards us. Schematic PV diagrams along and transverse
to the projected cone symmetry axis are shown in Fig.~\ref{f:cone_pv}. (Note that it is a very simple cone model with uniform density. The PV diagram would be much more complicated for a non-uniform cone, or multiple cones.) The resulting PV diagram along the main axis has two scenarios: one with constant outflow
velocity, and one where the gas is first accelerating and then decelerating. It is likely possible to find an outflow velocity scenario that can at least produce a reasonable
fit to the PV along the jet axis ((A) in Fig~\ref{f:pv}), if the northern cone is tilted towards us and the opening angle is large,  $>$$45^{\circ}$.  
This orientation of a wide angle cone is, however, inconsistent with the optical dust structure found by \citet{roussel06} (their Fig.~1) and \citet{heisler94} (their Fig.4c), which
would require the northern cone to be tilted away from us.  

\begin{figure}[tbh]
\includegraphics[scale=0.3]{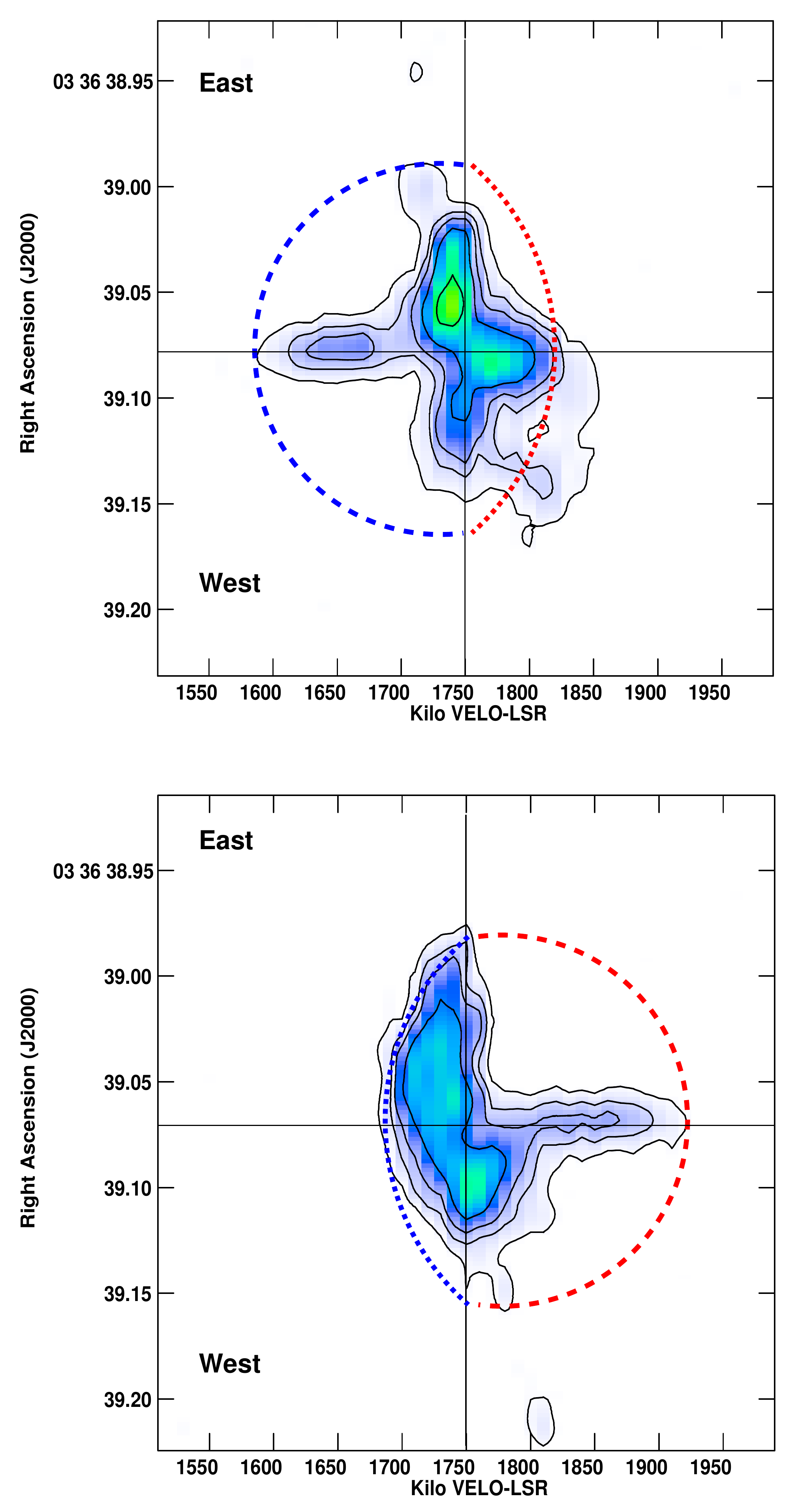}
\caption{\label{f:transverse} PV diagram, showing gas velocities in a slit across the jet axis at $\pm$0.\arcsec3. 
Contour levels are 3.1$\times$(1,2,4,8,16,32) mJy beam$^{-1}$. The colour scale ranges from -11 to 156 mJy beam$^{-1}$.
The dashed semi-ellipticals show the PV diagram expected from a projected wide-angle cone  It is clear that this model does not fit the data. 
}
\end{figure}

\smallskip
\noindent
Another important argument against the cone-projection model is the shape of the observed PV diagrams transverse to the jet symmetry axis (panels C- F in Fig.~\ref{f:pv}). For a cone, the PV diagrams perpendicular to the major axis will always be ellipses (see schematic PV diagrams in Fig.~\ref{f:cone_pv}).  And, when the top
cone is tilted towards us, there should be a broad (in space) blueshifted emission component to the north. 
The observed PV diagrams transverse to the jet axis (Fig.~\ref{f:pv}) do, however, not show this structure. In Fig.~\ref{f:transverse} we show the PV diagrams D and E from Fig.~\ref{f:pv} with the expected PV diagram of a tilted cone indicated by
dashed lines. Instead of tracing out the curved front ellipse, the maximum velocity is structured in a spatially unresolved tounge-like shape of broad emission.

\smallskip
\noindent
We note that this exercise is not an attempt to model the minor-axis structure of the low-velocity gas as discussed in Sect.~\ref{s:low}. The low-velocity gas
may (at least partially) originate in a cone-like slow outflow, which we suggest is interacting with the molecular jet. An attempt to link it to the optical dust structure
mentioned above would require its southern part to be at least slightly directed towards us.

 \end{appendix}

\end{document}